\documentclass[sn-apa]{sn-jnl}

\usepackage{amsmath}
\usepackage{amsfonts}
\usepackage{graphicx}%
\usepackage{multirow}%
\usepackage{amsmath,amssymb,amsfonts}%
\usepackage{amsthm}%
\usepackage{mathrsfs}%
\usepackage[title]{appendix}%
\usepackage{xcolor}%
\usepackage{textcomp}%
\usepackage{manyfoot}%
\usepackage{booktabs}%
\usepackage{algorithm}%
\usepackage{algorithmicx}%
\usepackage{algpseudocode}%
\usepackage{listings}%

\newcommand{\edgestate}{x}
\newcommand{\bumpstate}{y}
\newcommand{\edgeedgeweight}{w_r}
\newcommand{\edgeedgeweightunnormed}{\tilde w_r}
\newcommand{\edgebumpweight}{w_d}
\newcommand{\bumpedgeweight}{w_f}
\newcommand{\edgeendinput}{I}
\newcommand{\nonlinearity}{h}
\newcommand{\widthnormal}{\sigma_\nonlinearity}
\newcommand{\widthnarrow}{\sigma_r}
\newcommand{\feedback}{v}
\newcommand{\laplacefunction}{F}

\newcommand{\laplacefunctionalvar}{s}

\newcommand{\desiredlaplaceshape}{\laplacefunction_d}
\newcommand{\desirededgeshape}{x_d}
\newcommand{\genericfunction}{\mathcal{F}}

\begin{document}
\title[Continuous Laplace attractor networks]{Continuous Attractor Networks
for Laplace Neural Manifolds} 

\author*[1]{\fnm{Bryan C.} \sur{Daniels}}\email{bryan.daniels.1@asu.edu}

\author*[2]{\fnm{Marc W.} \sur{Howard}}\email{marc777@bu.edu}

\affil[1]{\orgdiv{School of Complex Adaptive Systems},
\orgname{Arizona State University}, \orgaddress{\street{PO Box 872701},
\city{Tempe}, \postcode{85287}, \state{AZ}, \country{USA}}}

\affil*[2]{\orgdiv{Department of Psychological and Brain Sciences},
\orgname{Boston University}, \orgaddress{\street{610 Commonwealth Ave},
\city{Boston}, \postcode{02215}, \state{MA}, \country{USA}}}

\abstract{
    Many cognitive models, including those for predicting the time of future
    events, can be mapped onto  a particular form of neural representation in
    which activity across a population of neurons is restricted to manifolds
	that specify the Laplace transform of functions of continuous variables. These
	populations coding Laplace transform are associated with another population that
	inverts the transform, approximating the original function. This paper presents
	a neural circuit
	that uses continuous attractor dynamics to represent the Laplace
	transform of a delta function evolving in time.  One population places an
	edge at any location along a \mbox{1-D} array of neurons; another population places
	a bump at a location corresponding to the edge.  Together these two populations can estimate a Laplace transform of delta functions in time along with an approximate inverse transform.  
 Building the circuit so the edge moves at an appropriate speed enables the network
 to represent events as a function of log time.  Choosing the connections appropriately
 within the edge network make the network states map onto Laplace transform with exponential
 change as a function of time.
  In this paper we model a learned temporal association in which one stimulus
  predicts another at some fixed delay \emph{T}.  Shortly after \emph{t}=0 the
  first stimulus recedes into the past. The Laplace Neural Manifold representing
  the past maintains the
  Laplace transform exp(\emph{-st}).  Another Laplace Neural Manifold represents the
  predicted future.  At \emph{t}=0, the second stimulus is represented a time \emph{T}  in
  the future.  At each moment between 0 and \emph{T}, firing over the Laplace transform
  predicting the future  changes as \mbox{exp[-\emph{s}(\emph{T}-\emph{t})]}. Despite exponential growth in
  firing, the circuit is robust to noise, making it a practical means to
  implement Laplace Neural Manifolds in populations of neurons for a variety of
  cognitive models.
  }

\maketitle

Many authors have proposed that the evolutionary adaptation of the brain is to
predict the future \citep{Clar13,FrisKieb09,Fris10,RaoBall99,PalmEtal15}.  Our
ability to predict the future depends on our ability to remember the past.  A
growing body of work has demonstrated that the nervous system of many animals
uses a Laplace transform, with real $s$, to construct a temporal memory of the
recent past \citep{AtanEtal23,TsaoEtal18,BrigEtal20,ZuoEtal23,CaoEtal24}.  
It has been hypothesized that populations of neurons representing the Laplace
transform of functions of time (and other variables) are paired with another
population that represents the inverse Laplace transform
\citep{ShanHowa10}. We refer to these paired representations of a
continuous function as a \emph{Laplace Neural Manifold}.

\begin{figure}
	\centering
   \includegraphics[width=0.7\columnwidth]{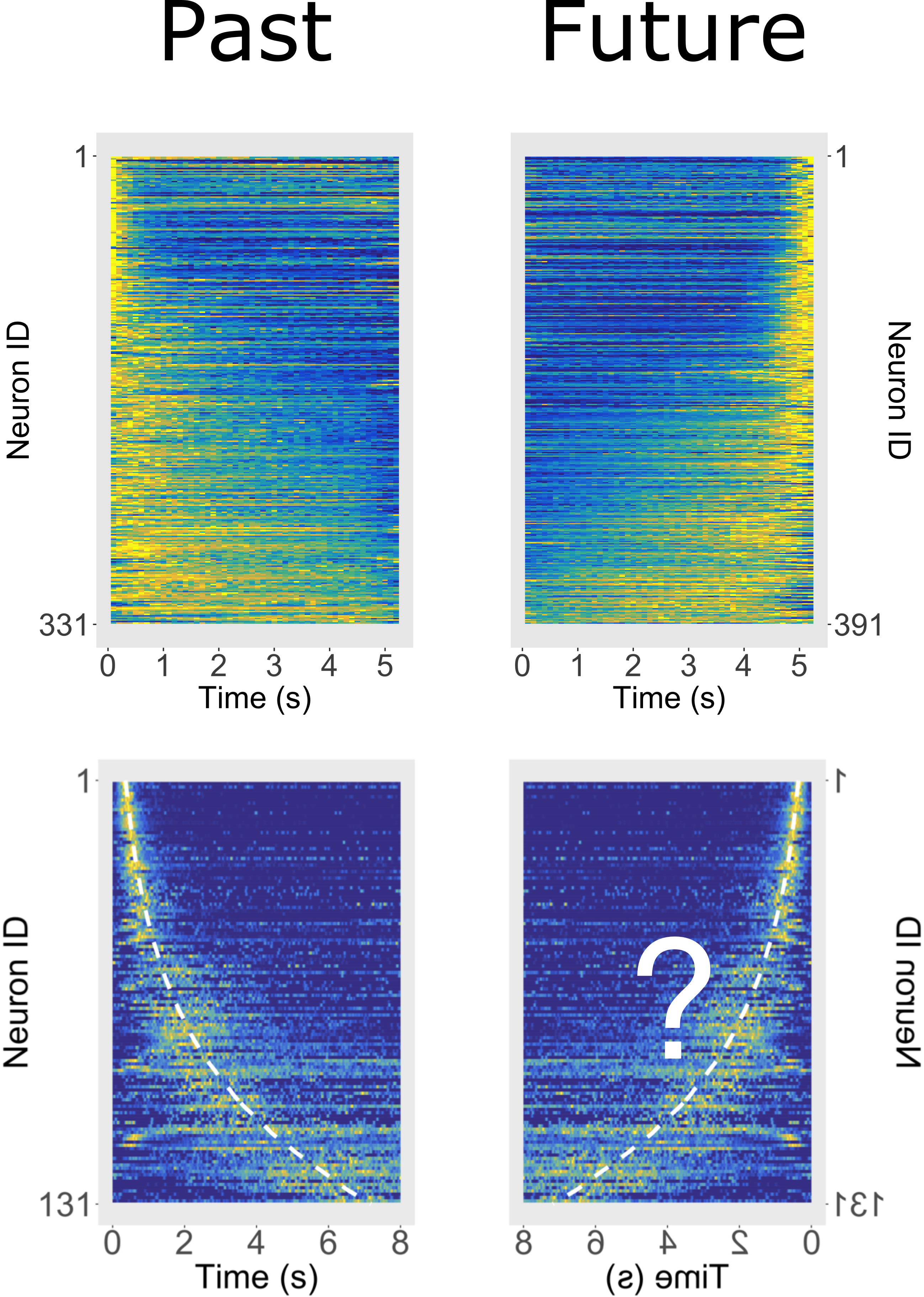}
	\caption{
    \textbf{Neural representations of the past (left) and future (right) in the brain.}
	\textbf{Top: Laplace transform of future and past in mPFC.} Recordings are
	during an interval reproduction task in which the animal reproduces an
	interval of duration $T$.  One group of cells decays exponentially $e^{-st}$
	with time during the reproduction phase (left).  These neurons peak at about
		the same time but have a continuous range of decay rates.
        This population 
		codes for the Laplace transform of a delta function at time $t$ in the past.
		Another population (right) ramps up as $e^{-s(T-t)}$ during the reproduction
		phase with a continuous range of
		ramp rates.  This group codes the Laplace transform of a delta function at time
		$T-t$ in the future. Data from Henke, et al., (2021).  Analyses by Cao, et
		al., (2024). 
  \textbf{Bottom: Inverse Laplace transform of the past in hippocampus.}
  Hippocampal time cells (left) fire in sequence following a task relevant
  stimulus.  The sequence is understandable as an approximation of the inverse
  Laplace transform of time since the sequence began.  After Cao, et al., 2023.
  Populations of neurons coding for the inverse of the future (right) would fire in in
  sequence approaching the predicted time of occurrence.  This population is
  hypothesized, so the figure is simply a reflection of the data on the left.
	\label{fig:DMmPFC}
		}
\end{figure}

In addition to memory for the past, it has also been proposed that the brain
uses a real Laplace transform to represent the time of predicted \emph{future} events
\citep{HowaEtal23}.  Some neural evidence suggests that this could be the case
\citep[Fig.~\ref{fig:DMmPFC}]{AffaEtal24,CaoEtal24}.  However,  constructing a
neural circuit that represents the Laplace transform of the time of a
predicted future event raises significant computational challenges that do
not arise in representing the remembered past.  Similar computational challenges arise in 
building realistic neural models for Laplace representations of 
spatial variables during allocentric navigation \cite{HowaEtal14} and
evidence accumulation \cite{HowaEtal18}, so that a solution to this problem could have
applications in many domains of computational cognitive neuroscience.

\begin{figure}
	\centering
	\includegraphics[width=0.8\columnwidth]{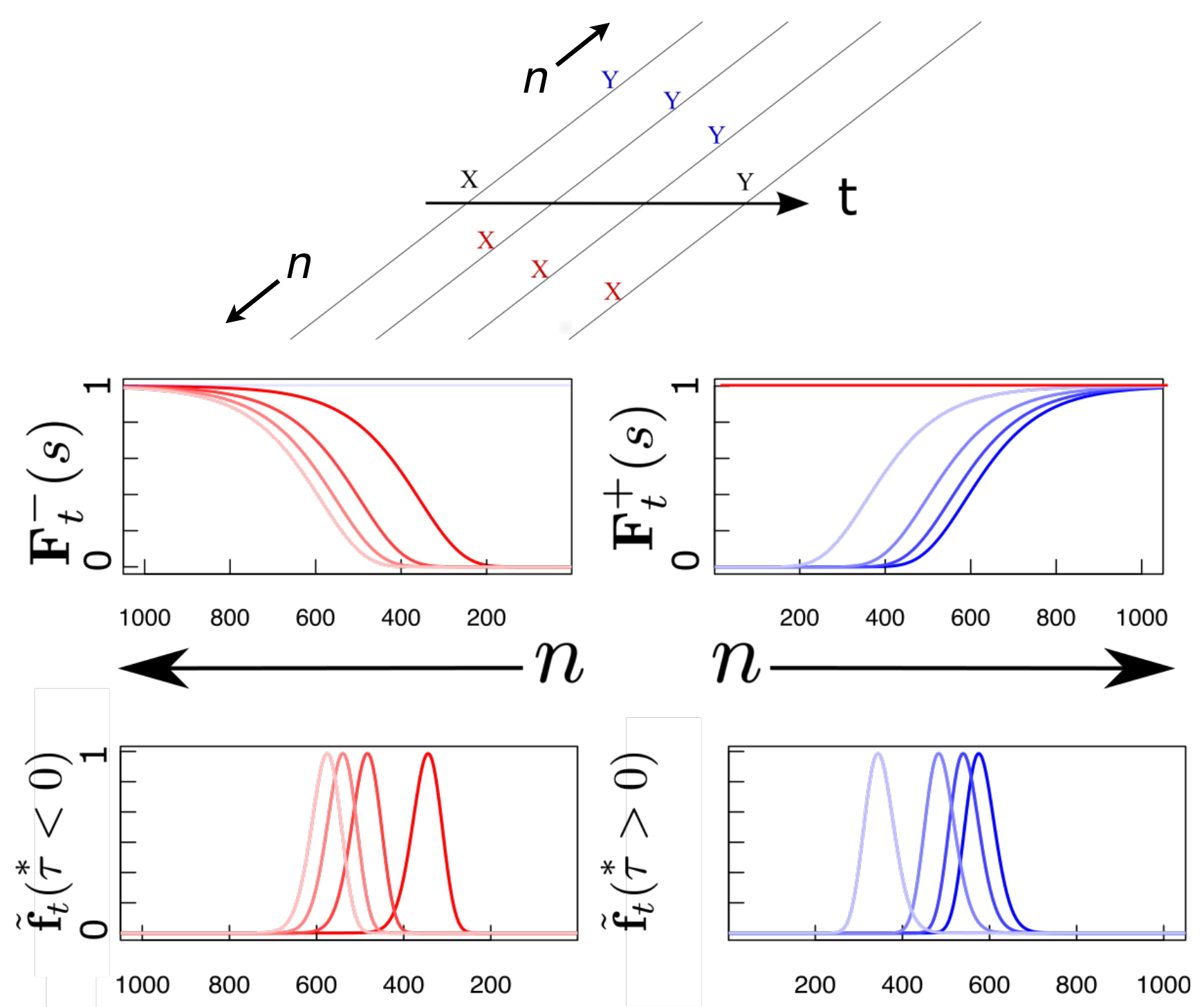}
	\caption{
		\textbf{Laplace Neural Manifolds 
            representing delta functions can be implemented using
		continuous attractor neural networks.}
  Top: Memory of the past and prediction of the future as a logarithmic timeline.
    Suppose that \textsc{x} and \textsc{y} are experienced consistently with a
    lag of $T$ seconds.  As time unfolds (horizontal line) after 
    presentation of \textsc{x} at $t=0$, the observer will have a memory for the past
    (red, below the line) and prediction of the future (blue above the line).  Because of the logarithmic compression of the timeline,
    the movement of the remembered/anticipated times do not change
    evenly as time passes.
        Middle: The function $e^{-s_n t}$ for
        evenly spaced choices of $\tau$ (red curves) plotted as a function of $n \propto \log s$.
        If we assume a log distribution of rate constants, such that $s_n=s_o a^n$ and $n\propto \log s$, then evolving $e^{-s_n t}$ 
        simply amounts to translation.  
        At time $t=0$, the future event is predicted a time $T$ in the future.  At time 
        $t<T$, the future event is predicted a time $T-t$ in the future.
        Evolving $e^{-s(T-t)}$ simply requires moving the edge in the opposite
        direction (blue curves). 
	   Bottom: A bump that translates as a function of time is 
        understandable as the inverse transform of the original function over $\log$ time convolved with some function of $\log$ time that describes the shape of the bump.
		\label{fig:bumps}
	}
\end{figure}

This paper addresses this challenge by demonstrating that continuous attractor
neural networks are well suited to construct and maintain a neural population
that represents the Laplace transform of future events.  This proposal
relies on two assumptions.  First, the network is restricted to represent the
Laplace transform of delta functions that can change in their location in time. 
Specifically, the Laplace transform of a delta function centered at $\tau$ is $e^{-s\tau}$.
Second, the values of $s$ across the population 
are distributed uniformly as a function of log time. 
That is, $s_n$, the rate constant for the $n$th neuron, scales as $a^n$ for some constant $a$
so that the change in $\log s$ for neighboring neurons is constant across all neurons.
This choice for the distributions is consistent with a large body of work in
psychology and neuroscience \citep{Fech60,VanEEtal84,GallGelm00,FeigEtal04}
and is consistent with empirical findings for time constants in at least some
brain regions \citep{CaoEtal23,GuoEtal21}.
When rate constants $s$ are chosen in this way, when $\tau$ changes to $c\tau$
the Laplace transform simply translates across the population of cells by an
amount $\log_{a} c$ (Figure~\ref{fig:bumps}).  
For a delta function, translation in time also maps onto rescaling in
time.  If we translate in time by taking $\tau \rightarrow \tau + b$, this is
equivalent to rescaling time $\tau \rightarrow c \tau$ with $c = \frac{\tau+b}{\tau}$.
Thus, the time evolution of Laplace transform of a delta function  can be
represented as translation across a continuous attractor network if the
rate of translation changes as time passes. 

One extremely simple way to compute the Laplace transform
of the past is to construct a bank of leaky integrators each with $s$ chosen
from a continuous range.  The dynamics for each integrator obey
\begin{equation}
    \frac{dF}{dt} = -s F + f(t).
    \label{eq:leakyintegrator}
\end{equation}
Let us use the notation $f_t(\tau)$ to refer the past leading up to the
present at time $t$ with $\tau=\infty$ referring to the distant past.
The solution of Equation~\ref{eq:leakyintegrator}  is
\begin{equation}
    F_t(s) = \int_0^{\infty} e^{-s\tau} f_t(\tau) d\tau
    \label{eq:wormFR}
\end{equation} from which we can see that Eq.~\ref{eq:leakyintegrator}, with
continuous values of $s$  across neurons describes a population that codes for
the real Laplace transform of the past.  If at some moment in time, the past
consists only of a delta function at one moment
$\tau$ in the past, then $F_t(s) = e^{-s\tau}$.  Suppose that the stimulus is
experienced at $t=0$.  As time passes, the stimulus will be $\tau=t$ in the
past and each neuron will decay
exponentially as $e^{-st}$.
Now consider a population of neurons coding for the Laplace transform of a
future event. Let us suppose that at $t=0$ the event is predicted $T$  seconds
in the future. At any moment $0<t<T$, the firing rate across the population
represents a time $\tau=T-t$.  As a function of time, the firing rate for
neurons representing the time of a future event should be
$e^{-s(T-t)}$. For this population, firing should ramp up exponentially as a
function of time with a continuous range of $s$.   Naturally the
exponential growth cannot continue indefinitely so at time $t=T$ we should
observe a discontinuous change as either the predicted stimulus is observed or
the prediction is violated.

The leaky integrators in Equation~\ref{eq:leakyintegrator} could be implemented
neurobiologically using one of many possible mechanisms to affect the functional
time constant of individual neurons.  
There is by now overwhelming evidence that there is a continuous diversity of
functional time constants in neurons within the same brain region across many
cortical and non-cortical regions
\citep{BernEtal11,CavaEtal20,SpitEtal20,DansEtal23,MauEtal18,CaoEtal22,BrigEtal20}.
Potential mechanisms include slow
intrinsic currents
\citep{FranEtal06,TigaEtal15}, network recurrence \citep{SterEtal23,DahmEtal19},
or synaptic mechanisms with a diversity of time constants
\citep{YoshEtal08,GuoEtal21,BarrEtal22}.   
In principle one could adapt Equation~\ref{eq:leakyintegrator} to code for the
future by initializing the population to $e^{-sT}$ then updating with
Equation~\ref{eq:leakyintegrator} but with  positive $s$.  However, this
approach is unacceptable.  Exponential growth implies that 
the population would be unstable in the presence of any amount of noise.

\section{Continuous attractor model for Laplace/inverse representations }
The primary contribution of this paper is demonstrating that continuous
attractor networks can be used to evolve $F_t(s)$ in time if two conditions
are met. First, at any particular moment we only need to represent the
Laplace transform of a delta function.  Second,  the values of $s_n$ are in a
geometric series.   If these two conditions are met, then evolving $F_t(s)$ in
time amounts to translating an edge of differential activation across the
population.   Continuous attractor neural networks are well-suited to this
problem.

To see why these conditions are sufficient, consider how the solution to
Equation~\ref{eq:leakyintegrator} changes between time $t$ and $t+\Delta t$
(Figure~\ref{fig:bumps}).  In general, $F_{t+\Delta t}(s) = e^{-s \Delta t}
F_t(s)$.  Let the population code a delta function at time $\tau$ in the past
so that $F_t(s_n)=e^{-s_n\tau}$.  Now, $F_{t+\Delta t}(s_{n+\Delta n}) =
e^{-s_{n+\Delta n}(\tau+\Delta t)}$; if we want the time evolution to be a
simple translation by $\Delta n$, we require that $\frac{s_{n+\Delta n}}{s_n}
=\frac{\tau}{\tau+\Delta t}$. 
If $\Delta n$ is a
constant as a function of $n$, then we can say that the activity has
translated across the population with the passage of time. $\Delta n$ is a
constant iff the $s$ are in geometric series.  Note that the magnitude of the
translation along $n$ depends not only on $\Delta t$, but on $\tau$ as well.
Defining the ratio between adjacent $s$  as
$a = \frac{s_{n+1}}{s_n}$, we find that 
$
	\Delta n = \log_a \frac{\tau}{\tau + \Delta t}
$.

The continuous attractor network proposed here  consists of two populations.
The population that represents the Laplace transform is an attractor network
with interactions that favor neighbors in the same state of activation, and
with the two ends of the array clamped to different values.  This network
exhibits a stable ``edge'' at any point along the array. If the interactions
are chosen to be strong enough (as in a ferromagnet at low temperature), this
network maps precisely onto the Laplace transform of a delta function with
values of the rate constant chosen in a geometric series. The ``edge
attractor'' network representing the Laplace transform is coupled to a
``bump'' network that takes as input the derivative of the edge attractor
network, placing the bump at a location that corresponds to the location of
the edge. With appropriate feedback coupling between the networks, the bump
network pushes the edge along at a pace that enables the location of the edge
to be mapped onto $\log \tau$.  

Consider the activity of an individual cell at a particular location in the
middle in the bump network. Initially, the cell's firing is low. As the bump
approaches the cell's location,  its firing increases.   As the bump moves past
the cell's location its firing rate decreases again.  As the bump moves monotonically
from one end of the network to the other, each cell is activated in this way at a time that
depends on their location in the network and the rate at which the bump moves.
These sequentially activated cells have behavior that resembles that of ``time cells''
(Figure~\ref{fig:DMmPFC}, bottom left). Time cells were initially reported in
hippocampus \citep{PastEtal08,MacDEtal11,CaoEtal23} and subsequently reported in
many other brain regions \citep{TigaEtal18a,JinEtal09,MellEtal15}. If time cells
result from movement of a bump of activity across a continuous attractor network
then there is no mechanistic obstacle to constructing ``future time cells''
(Figure~\ref{fig:DMmPFC}, bottom right) by simply having the bump travel in the
opposite direction.   These ``future time cells'' have not been characterized in
the hippocampus (although spatial predictive sequences have been observed
\citep{FerbShap03,SareEtal17,GautTank18}). There is some evidence for predictive
sequences in other brain regions, including cerebellum \citep{WagnEtal17},
although the empirical story is not nearly as clear as for standard time cells.

The goal of this model is to develop a representation of what happened when in
the past, and of what will happen in the future.  
The requirement that the input be a delta function places strong constraints
on the ``what'' information that can be represented by this circuit.  
This assumption seems reasonable to make sense of working memory experiments
where the participant is presented with a small number of discrete symbols
chosen from a larger set (for instance, letters or
images).  To construct a multi-item working memory we could imagine that these
symbols are each associated with  distinct timelines corresponding to
edge/bump attractors extending from the past into the future.   
One could imagine that presentation of the stimulus causes a delta
function input corresponding to its onset. 
The expected time of future events is in general not a delta function.  One can
imagine that the future timeline consists of a delta function sampled from the
true distribution \cite{HowaEtal24}.
More elaborate representations are conceivable, but in any event, the
difficulty of representing Laplace transforms of arbitrary functions seems
like a fundamental constraint 
for predicting the time of future events.

\subsection{Overview of the circuit model}

To generate the Laplace transform of a delta function using neurons with logarithmically
distributed rate constants we require 1) a continuous attractor neural network
with an edge attractor with the correct shape and 2) translation of the edge over
the network in time with a speed $d\bar n/dt$ proportional to $1/\tau$.  
Inverting the transform can be accomplished with another continuous attractor
neural network that places a bump of activity in the same location as the
edge.  Because the location of the edge is understandable as log time, so too
the bump also translates as a function of log time.  Because the bump should
have the same
shape at any moment in the past, neurons in the bump network will naturally
inherit scale-invariant properties observed for hippocampal time cells
\citep{CaoEtal22}.  Because the bump has the same location 
as the edge, the pattern over the bump network is understandable as the original
function over log time, convolved with a function of log time that describes the
shape of the bump. In this sense, the state over the bump network approximates
the inverse Laplace transform, but as a function of logarithmic time.

\begin{figure}
    \centering
    \includegraphics[width=0.5\columnwidth]{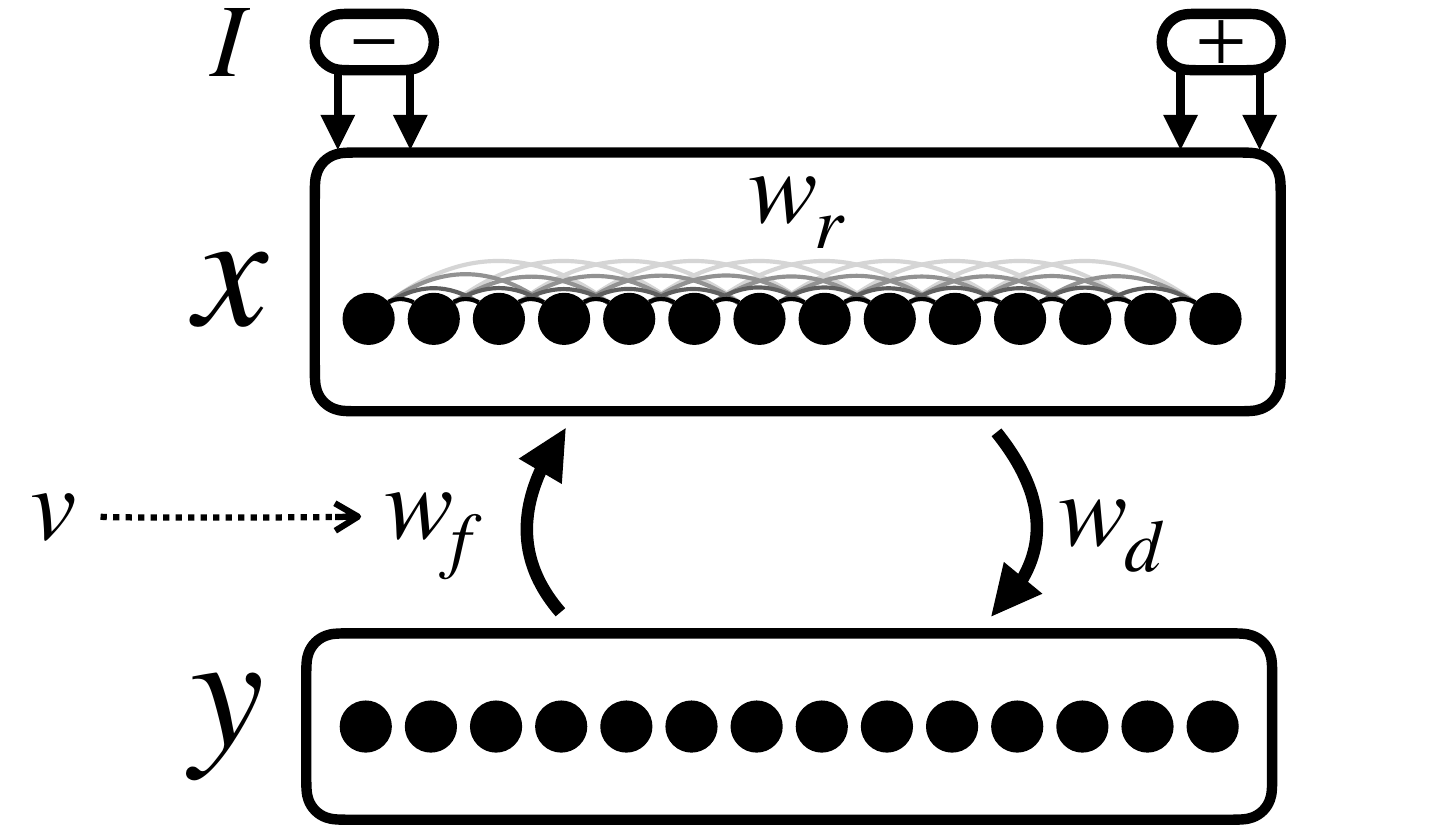}
    \caption{\textbf{Schematic of network setup.}   
    The edge attractor is distributed over one set of units $\edgestate(n)$, utilizing 
    recurrent connections $\edgeedgeweight$ within this set of units.  Another
	set of units $\bumpstate(n)$ supports a bump.  Connections
	$\edgebumpweight$ from $\edgestate$ provide input to $\bumpstate$.
	These connections estimate the derivative with respect to $n$.
	Connections from $\edgestate$ to $\bumpstate$, $\bumpedgeweight$, move
	the location of the edge as a function of time.  Because the strengths
	of these connections are a function of $n$, the velocity of the edge
	$\feedback$ depends on the location of the edge.  
	\label{fig:network-schematic} }
\end{figure}

We want an edge attractor to approximate $\mathcal{L}\{ f\}(s)$ for delta functions
$f(t-\tau)$, with real values of $s$ sampled along a log scale.  
Our strategy will be to use two sets of neurons
(Fig.~\ref{fig:network-schematic}) labeled by index $n$.
We will treat $n$ as continuous and align the two populations with one another.
We will design the connections within and between these two
sets of neurons $\edgestate(n)$ and $\bumpstate(n)$  so that their stable states
are simply related to the Laplace transform of $f$ and the inverse transform
respectively.   This paper builds a minimal circuit to accomplish these goals and does not 
attempt a computationally optimal nor biologically detailed model.

The real Laplace transform of a delta function at a point on the positive line
takes values between zero and one. The continuous attractor neural network
$\edgestate(n)$ goes from $-\edgestate_\mathrm{max}$ to
$\edgestate_\mathrm{max}$.  
We provide external input to clamp the two edges of the network at $n=1$ and
$n=N$ to these extreme values.  Interactions
between units within
$\edgestate(n)$ encourage neighboring units to have the same activation.  
These interactions are implemented by connections $\edgeedgeweight$;
the weights on the connections between units within $\edgestate$ depend only on their distance in
the network, $n-n'$. 
If the parameters of the network are chosen appropriately (i.e., with low effective temperature), then
somewhere between the ends of the network, there should be
a transition between maximal and minimal values. 
This boundary, or edge,
can be at any location along the network
as long as the range of the local interactions is far from the ends $n=1$ and $n=N$.
The requirement that the neurons code real Laplace transform requires a specific
shape of the edge, which in turn requires a specific form for connection weights.

In this paper we adopt a minimal approach to the bump network.  Rather than
recurrent connections within $\bumpstate(n)$ we simply take input to neurons in
$\bumpstate(n)$ to be large if there is an edge in the neighborhood of each $n$.

The input to $\bumpstate(n)$  from $\edgestate(n)$ is provided by
$\edgebumpweight$, which simply computes the derivative of $\edgestate(n)$ in
the neighborhood of $n$.  Again, the connections between $\edgestate(n)$ and
$\bumpstate(n')$ depend only on $n-n'$. The bump then provides
feedback to the $\edgestate(n)$ via connections $\feedback(n)$, causing the edge
to move over time.  By choosing $\feedback(n)$ to depend on $n$ appropriately we
can control the rate at which the edge moves at different points along the network.

\subsection{Continuous attractor neural network with edge solutions }
To define the neural dynamics, 
we use a simple neural firing rate model:
\begin{eqnarray}
    \label{eq:edge_ode}
    \frac{d \edgestate(n)}{dt} &=& 
        -\edgestate(n) 
        + \edgeendinput(n)
        + \sum_{n'} \edgeedgeweight(n-n') \nonlinearity \left[ \frac{\edgestate(n')}{\widthnarrow} \right]
            + \bumpedgeweight(n-n') \nonlinearity \left[\frac{\bumpstate(n')}{\widthnormal} \right]
        + \xi \\
    \label{eq:bump_ode}
    \frac{d \bumpstate(n)}{dt} &=& 
        -\bumpstate(n) 
        + \sum_{n'} 
            \edgebumpweight(n-n') \nonlinearity \left[\frac{\edgestate(n')}{\widthnormal}\right]
        + \xi.
\end{eqnarray}
Here $\edgestate$ and $\bumpstate$ represent instantaneous relative neural firing rates.  $\edgeendinput(n)$ consists of a large negative input 
current $-i$ applied from $n=1$ to $n_i$ and large positive input current $+i$ applied at the other
end of the network from $n=N$ to $N-n_i$, which sets boundary conditions so that the edge has
an orientation with small $n$ neurons
saturated at $-\edgestate_\mathrm{max}$ and large $n$ neurons saturated at
$+\edgestate_\mathrm{max}$. 
The kernels
$\edgeedgeweight$, $\edgebumpweight$, and $\bumpedgeweight$ define the strengths of interactions
among neurons and are derived explicitly in section~\ref{sec:kernels}.
The kernel $\edgeedgeweight$ defines connections
among $\edgestate$ neurons with varying $n$ and determines the shape of the
edge. The kernel $\edgebumpweight$ defines connections from $\edgestate$ to
$\bumpstate$, and is set to implement an effective derivative of 
$\edgestate(n)$ being represented in $\bumpstate(n)$.
The kernel $\bumpedgeweight$ defines feedback from $\bumpstate$ neurons to
$\edgestate$ neurons. 
The nonlinear function
$\nonlinearity = \tanh$ 
defines the mapping from neural state
to synaptic current, with width $\widthnarrow$ among $\edgestate$ neurons,  
and $\widthnormal$ for all other synaptic connections.
Uncorrelated Gaussian noise is represented by $\xi$, parameterized
such that variance is added at a rate $\sigma^2_\xi$.

In simulations, we initialize the neural states to a stationary state
$\edgestate^*(n)$, $\bumpstate^*(n)$, found numerically by setting feedback
$\bumpedgeweight$ and noise $\sigma_\xi$ to zero, with the edge and bump set
initially at location $\bar n_0$.
Translation equivariance of $\edgeedgeweight$, $\edgebumpweight$, and
$\bumpedgeweight$ implies that the edge can smoothly translate to any location
$\bar n$ along the network.  In implementing cognitive functions, one might
imagine using input to set the initial location of the edge.
Once the network is in a stable edge state, this input can be removed.  

\subsection{Mapping edge solutions to the real Laplace transform}
We desire that there be a mapping from neuron index $n$ to the Laplace variable
$\laplacefunctionalvar$, $\laplacefunctionalvar(n)$.
At any moment, the Laplace transform of a delta
function at time $\tau$ gives  $F(n) = e^{-s(n) \tau}$. 
Therefore, if the network has successfully represented Laplace transform of a
delta function, we should be able to observe a particular unit as time evolves
and see exponential changes in firing with rate constant $s(n)$.  
To ensure the time evolution of the network is simply translation of the edge,
we require that $s(n) \propto a^n$ for some constant $a$.  Equivalently, we
require $n = \log_a \frac{s}{s(n=1)}$, so that the $s(n)$ form a logarithmic
time scale for location of the edge.  
Thus we set the size of the scale factor between $s$ of neighboring neurons to be $a$, with $a < 1$.
In the limit that $a \rightarrow 1$, we get a continuum limit
in which neighboring neurons represent infinitesimally distinct $s$.  

The dynamics defined by Equation~\ref{eq:edge_ode} 
produce $\edgestate$ varying between $-\edgestate_\mathrm{max}$ and
$+\edgestate_\mathrm{max}$.  The Laplace transform of a delta function with
$\tau > 0$ should vary between 0 and 1, so we define
\begin{eqnarray}
    \label{eq:laplacemapping}
    \laplacefunction(\laplacefunctionalvar) &=&  \frac{1}{2} \left[ \frac{\edgestate(\laplacefunctionalvar)}{\edgestate_\mathrm{max}} + 1 \right] .
\end{eqnarray}
To get the stable states as a function of $n$ to  be identified with Laplace
transform of a delta function, we need the edge to have the right shape 
and for the edge to move at the proper speed. Given that those problems have been
solved (see below), the activity over $\bumpstate(n)$ will map onto the inverse
Laplace transform as long as the bump is located at the same $n$ and does
not change shape as the edge moves along $n$.   If these conditions are met,
then the bump state is understandable as the convolution of the delta function
$f$ as a function of $\log \tau$ and some function of $\log \tau$ that describes the shape 
of the bump over $n$.   

\subsection{Setting the interaction kernels}
\label{sec:kernels}
We will first set aside the problem of defining $\edgeedgeweight$, which sets
the equilibrium shape of the edge, and merely assume that $\edgestate(n)$ is
monotonically increasing in $n$, with an edge location defined as the
(interpolated) $\bar n$ at which $\edgestate(\bar n) = 0$.
To get such an edge state to move, we will first approximate the derivative 
of $\edgestate$ with respect to $n$ in the states of
the $\bumpstate$ neurons, then use $\bumpstate(n)$ to 
drive the edge at the desired speed.

\subsubsection{Forming a bump in $\bumpstate$}
First, we want $\bumpstate(n)$ to represent a (discrete) derivative of $\edgestate(n)$:
\begin{equation}
    \label{eq:edge_derivative}
    \bumpstate(n) \rightarrow \frac{1}{2} \left[ \edgestate\left(n+1\right) -
	\edgestate\left(n-1\right) \right].
\end{equation}
In equilibrium assuming constant input and neglecting noise, from Eq.~\ref{eq:bump_ode}, 
\begin{equation}
    \bumpstate(n) \rightarrow \sum_{n'} 
            \edgebumpweight\left(n-n'\right) \nonlinearity\left[\edgestate\left(n'\right) / \widthnormal \right].
\end{equation}
If we set the scale of the nonlinearity $\widthnormal$ large enough, then
$\nonlinearity\left[\edgestate\left(n'\right)\right/\widthnormal] \approx \edgestate(n')/\widthnormal$,
and we can implement the derivative simply by setting
\begin{equation}
    \edgebumpweight(n-n') = \frac{\widthnormal}{2}\left[
	    \delta\left(n-n'-1\right) - \delta\left(n-n'+1\right) \right].
\end{equation}

\subsubsection{Moving the edge to form a logarithmic time scale}
With neural states $\bumpstate(n)$ representing the derivative of the edge as
a function of $n$ (Eq.~\ref{eq:edge_derivative}), this can be rescaled and
applied as an input to the $\edgestate$ neurons to move the edge.  We then
define $\bumpedgeweight$ to feed $\bumpstate$ back to $\edgestate$ with
a scaling factor $\feedback$ that can depend on $n$:
\begin{equation}
    \bumpedgeweight(n-n') = -\widthnormal \feedback(n) \, \delta(n-n').
\end{equation}
A constant feedback strength $\feedback(n) = \feedback_c$ moves the edge with constant speed $\frac{d \bar n}{dt} = \feedback_c$.  To see this, consider solving for the input necessary to move an equilibrium edge with shape $\edgestate^*(n)$ a small distance $\Delta \bar n$ in time $\Delta t$.  Since there is a continuous set of steady state edges with shape $\edgestate^*(n)$, the needed change in $\edgestate$ as a function of $n$ is $\Delta \edgestate = \edgestate^*(n - \Delta \bar n) - \edgestate^*(n) \approx -\Delta \bar n \frac{d\edgestate^*(n)}{dn}$.  Then near equilibrium the necessary input term added to $\frac{d\edgestate}{dt}$ to produce motion with speed $v_c = \frac{\Delta \bar n}{\Delta t}$ is $ -v_c \frac{d\edgestate^*(n)}{dn}$.  Finally, with $\bumpstate(n) \approx d\edgestate^*(n)/dn$, $\nonlinearity[\bumpstate(n')/\widthnormal] \approx \widthnormal d\edgestate^*(n)/dn$, so we need $\bumpedgeweight = -\widthnormal \feedback_c \delta(n-n')$.   

Instead of having the edge move with constant speed, we want $\bar n$ to grow
logarithmically in time:
\begin{equation}
    \bar n(t) = \bar n_0 + \log_a{\frac{t}{t_0}},
    \label{eq:barntime}
\end{equation}
corresponding to edge speed 
\begin{equation}
    \frac{d \bar n}{dt} = \frac{1}{\ln{a}} \frac{1}{t}.
    \label{eq:desired_velocity_vs_t}
\end{equation}As a function of $\bar n$, this corresponds to 
\begin{equation}
    \frac{d \bar
n}{dt} = \frac{1}{\ln{a}} \frac{1}{t_0} a^{-(\bar n - \bar n_0)}
    \label{eq:desired_velocity_vs_n}
\end{equation}
because $t(\bar n) = t_0 a^{(\bar n - \bar n_0)} $.  

If we make the assumption that the bump is sufficiently localized in $n$ such that we can replace variations in $\feedback(n)$ along the bump with the value at the edge location $\feedback(\bar n)$, then we get the desired speed by setting 
\begin{equation}\feedback(n) = \feedback_0 \, a^{-(n - \bar n_0)}
\label{eq:velocity}
\end{equation}
where we have replaced $\bar n$ by $n$ in the desired $\frac{d \bar n}{dt}$
and set the initial speed $ \feedback_0 = \frac{1}{t_0} \frac{1}{\ln{a}}$.

In our simulations, we must limit the size of $\feedback$ to avoid an instability that arises due to the feedback loop from edge to bump to edge neurons.  
We therefore limit $\feedback$ to a predefined $\feedback_\mathrm{max}$. 
This does not affect the edge motion as long as the range of $n$ over which the limitation acts does not significantly overlap the edge locations.  

\subsubsection{Reversing velocity of the edge and bump to represent events as a
function of future time}

To change the direction in which the edge moves, we can simply flip the sign of the desired
velocity $\feedback(n)$. Note that the dependence of the initial velocity $v_0$
on $t_0$ means that the edge moves in the correct direction when $t_0 > 0$
corresponds to representing past events, with the edge moving toward larger $n$,
and $t_0 < 0$ corresponds to representing future events, with the edge moving
toward smaller $n$.
If at time $t=0$ the future event is predicted at future time $T$, then 
at time $0<t<T$, the $\edgestate(n)$ corresponds to Laplace transform of 
a delta function at $\tau=T-t$.  In parallel, the network representing the past
represents a delta function at $\tau=t$.

In our example simulations, we demonstrate the representation of an event in the
past (at $\tau = 0$) that is going further in the past by setting $t_0 = 25$ and
letting time $t$ run forward to $t = 175$.  We demonstrate the representation of
an event predicted to happen in the future (also at $\tau = 0$) by setting $t_0
= -175$ and letting time $t$ run forward to $t = -25$.

\subsection{Setting the shape of the edge to ensure time course of each unit
in $\edgestate(n)$ is exponential}
Finally, we set the edge shape by solving for the form of interactions $\edgeedgeweight$ among $\edgestate$ neurons.  Note that an edge of any shape that moves with speed $\propto 1/t$ corresponds to $\edgestate$ decaying in time for a given $n$ at a rate $\laplacefunctionalvar(n)$ that decreases geometrically with $n$: 
\begin{equation}
    \laplacefunctionalvar(n) = \frac{1}{t_0 \, a^{(n - \bar n_0)}}.
\end{equation} 
Here we set the constant prefactor by stipulating that $\laplacefunctionalvar(n
= \bar n(t)) = t^{-1}$ for all $t$.  Specifically, the neuron corresponding to
the starting location $n = \bar n_0$ has rate constant $\laplacefunctionalvar =
t_0^{-1}$. 
The shape of the decay of a given neuron's state with respect to time will be some function
\begin{equation}
    \edgestate(t,n) = \genericfunction \left[\laplacefunctionalvar(n) \, t\right].
    \label{eq:scaling-function}
\end{equation}
$\genericfunction(\laplacefunctionalvar \, t)$ will be set by the shape of the edge as a function of $n$, which is in turn controlled by $\edgeedgeweight$.  
For $\edgestate(n)$ to correctly represent the Laplace transform $\laplacefunction(\laplacefunctionalvar)$ when rescaled as in Eq.~\ref{eq:laplacemapping},
we want the state of individual neurons to decrease exponentially as a function of time.
As a function of $t$ and $n$, we want the output to be
\begin{equation}
    e^{-A \, \laplacefunctionalvar(n) \, t} = \exp{\left[-A\frac{t}{t_0} a^{-(n-\bar n_0)}\right]},
\end{equation}
with a constant $A$ that we set to $A = \ln 2$ to fix the edge location $\bar n = \bar n_0$ at time $t = t_0$.  
At a fixed time $t$ with corresponding edge location $\bar n$ (setting $t = t_0$ and $\bar n = \bar n_0$), this corresponds to a desired functional shape of
\begin{equation}
    \desiredlaplaceshape(n,\bar n) = \exp{\left[-A a^{-\left(n-\bar
	n\right)}\right]}.
\end{equation}
In terms of neural states, we need, from Equation~\ref{eq:laplacemapping}
\begin{equation}
    \label{eq:desired_rates}
    \frac{\desirededgeshape(n,\bar n)}{\edgestate_\mathrm{max}} =  2 \desiredlaplaceshape(n, \bar n) - 1.
\end{equation}


The equilibrium shape of the edge as a function of $n$ is the solution of  
\begin{equation}
    \label{eq:equilibrium-shape}
    \edgestate^*(n) = \sum_{n'} \edgeedgeweight(n-n')
	\nonlinearity \left[\edgestate^*\left(n'\right) / \widthnarrow \right].
\end{equation}
using Eq.~\ref{eq:edge_ode} and neglecting terms corresponding to feedback,
noise, and inputs at the boundaries.
We then want to solve Eq.~\ref{eq:equilibrium-shape} for $\edgeedgeweight$ given 
$\edgestate^*(n) = \desirededgeshape(n)$ as in Eq.~\ref{eq:desired_rates}.

Note that, in the limit $\widthnarrow \rightarrow 0$, the $\tanh$ nonlinearity becomes
a step function.  In this limit, and assuming $x^*(n)$ has a single zero
crossing at $\bar n$, Eq.~\ref{eq:equilibrium-shape} produces
\begin{equation}
    \edgestate^*(n+1) - \edgestate^*(n) = -2 \edgeedgeweight(n - \bar n).
\end{equation}
Thus we can get the desired edge shape by setting the interaction kernel proportional to the derivative of the desired edge shape with respect to $n$ \citep{Amar77}:
\begin{equation}
    \edgeedgeweight(n - n') 
    \propto - \frac{d\desiredlaplaceshape(n,n')}{dn} 
    \propto a^{ -(n-n')} \desiredlaplaceshape(n,n') .
\end{equation}
The overall scale factor of the interaction kernel sets the scale of the
equilibrium neural states, $\edgestate_\mathrm{max}$. To set
$\edgestate_\mathrm{max}$ precisely in the discrete case, we numerically
normalize the interaction kernel, resulting in 
\begin{equation}
    \edgeedgeweight(\Delta n) = \edgestate_\mathrm{max} \frac{\edgeedgeweightunnormed(\Delta n)}{\sum_{\Delta n} \edgeedgeweightunnormed(\Delta n)},
    \label{eq:edgeedgeweightresult}
\end{equation}
where we define $\Delta n = n - n'$ and
\begin{equation}
    \edgeedgeweightunnormed(\Delta n) = a^{-\Delta n} \exp{[ -A \, a^{-\Delta n} ]}.
    \label{eq:asymmetric-kernel}
\end{equation}
As can be seen from Eq.~\ref{eq:asymmetric-kernel}, these weights are
asymmetric (they are not even in $\Delta n$).  The network will give an edge for many choices of weights; this
expression, coupled with the velocity of the edge as specified in
Eq.~\ref{eq:velocity}, gives exponential decay in time.

\subsubsection{Simulation methods}

\begin{table}
    \centering
    \begin{tabular}{l|l}
        $N$ & 100 \\
        $n_0$ & 50 \\
        $\widthnormal$ & 20 \\
        $\widthnarrow$ & 0.5 \\
        $a$ & $e^{0.25}$ \\
        $\edgestate_\mathrm{max}$ & 2 \\
        $\feedback_\mathrm{max}$ & 1 \\
        $i$ & 100 \\
        $n_i$ & 5 \\
        $\Delta t$ & $10^{-1}$ \\
        $\sigma_\xi^2$ & $10^{-3}$ \\
        $\delta n$ & $0.1256$ \\
    \end{tabular}
    \caption{Simulation parameters.}
    \label{tab:parameters}
\end{table}

Here we illustrate properties of the continuous attractor neural network model
with simulations, integrating the dynamics using a simple Euler timestep $\Delta t$.
In our simulation, setting $\widthnarrow = 0$, i.e., using a step
function nonlinearity instead of $\tanh$, does successfully create the desired
edge shape, but the
infinitely sharp interactions effectively pin the edge to the lattice, not
allowing the edge to move smoothly along $n$.  That is, the continuous nature of
the attractor states is interfered with by the discreteness of the lattice.  
To
mitigate this effect while still keeping close to the desired edge shape, we
reduce $\widthnarrow$ as much as possible before pinning becomes strong.  We find that
$\widthnarrow = 0.5$ is a good compromise, and that values between $0.15$ and $0.5$
produce relatively indistinguishable results.

A second complication due to discretization comes from the fact that setting
$\edgeedgeweight$ using the above logic, which assumes continuity, on discrete
lattices leads to an edge that moves slowly over time even with zero input
feedback.  To fix this, we numerically solve for an offset value $\delta n$ that
produces zero change in the edge's location after one application of the offset
interaction kernel $\edgeedgeweight(n-n'+\delta n)$.  Given our other simulation
values, we find $\delta n = 0.1256$.

\section{Results}

\begin{figure}
    \centering
    \includegraphics[width=\columnwidth]{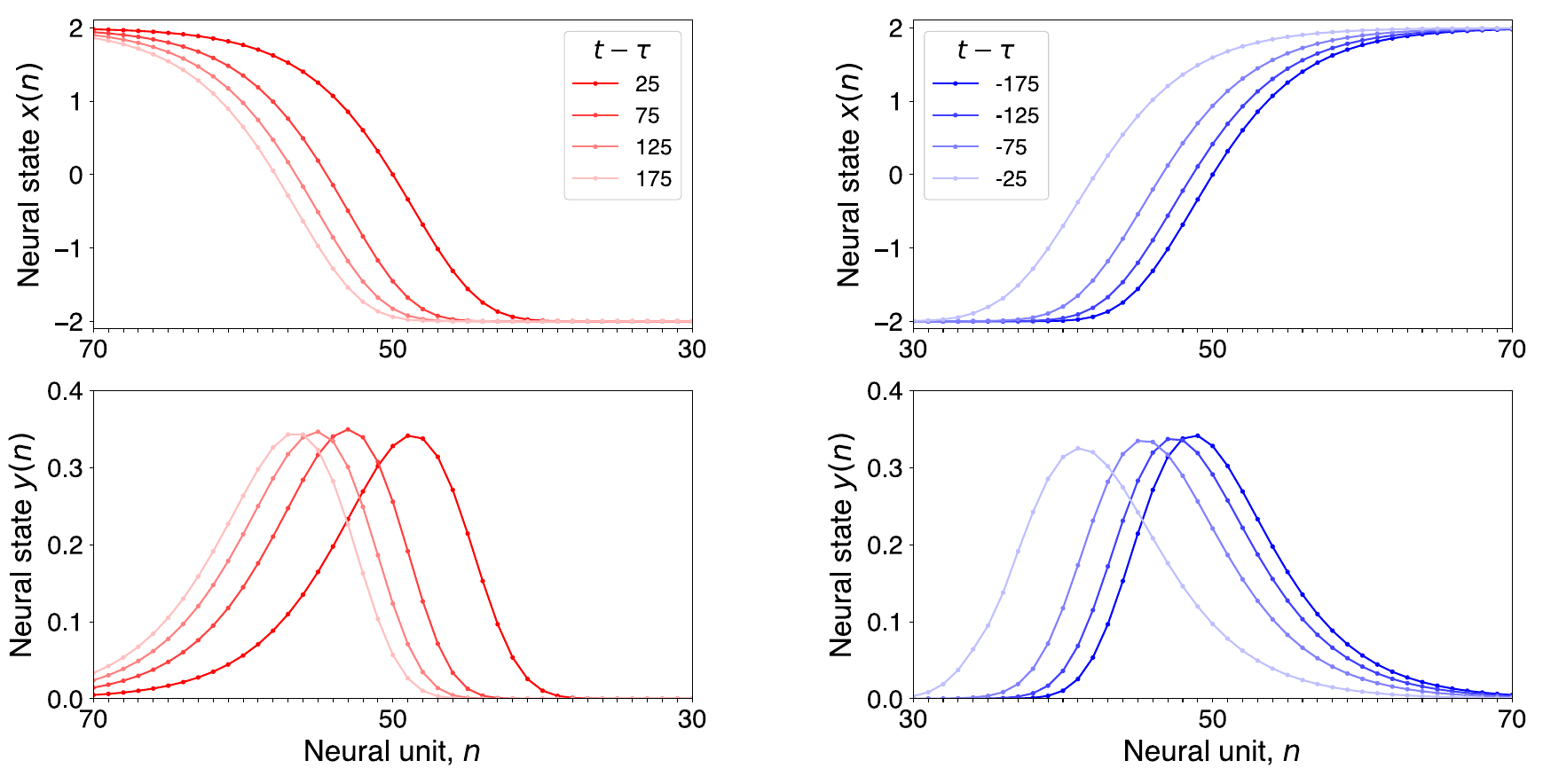}
    \caption{\textbf{Simulated dynamics of  continuous attractor neural network model for Laplace
    transform  of events in the past and the future.} 
    Layout is as in Figure~\ref{fig:bumps}.  Top: Activation $\edgestate(n)$
    over the network at four different time points.  
    Bottom: Activation $\bumpstate(n)$ at the same time points.
    Left: Networks representing the past $e^{-s\tau}$ at evenly-spaced $\tau$.
    Note that the edges become closer together as events recede into the past.
    Right:  Networks representing the future $e^{-s\left(T-\tau\right)}$ at
    evenly-spaced $\tau$.   Note that the edges become
    more widely separated as the event approaches from the future. 
    Here noise $\sigma_\xi^2 = 0$, with other simulation parameters listed in Table~\ref{tab:parameters}
    \label{fig:asymmetric-translation}
    }
\end{figure}

Figure~\ref{fig:asymmetric-translation} shows the activity over the network at
different evenly-spaced moments in time. At $t=0$, a stimulus that predicts a
future event at time $T$ is presented.  At time $t>0$ neurons in the
$\edgestate(n)$ mapping onto the Laplace transform of the past represent a time
$\tau=t$ in the past; neurons in the $\edgestate(n)$ mapping onto Laplace
transform of the future represent a time $\tau=T-t$ in the future. The
simulation of the continuous attractor network exhibits the desired properties
and gives results closely analogous to Figure~\ref{fig:bumps}. With the passage
of time the edge/bump changes its position but retains its shape. When  $\tau$
is bigger, the edge/bump complex is further from $n=1$.  As time passes, the
edge/bump representing the past moves more slowly whereas the edge/bump
representing the future moves more rapidly.

\begin{figure}
    \centering
    \includegraphics[width=0.8\columnwidth]{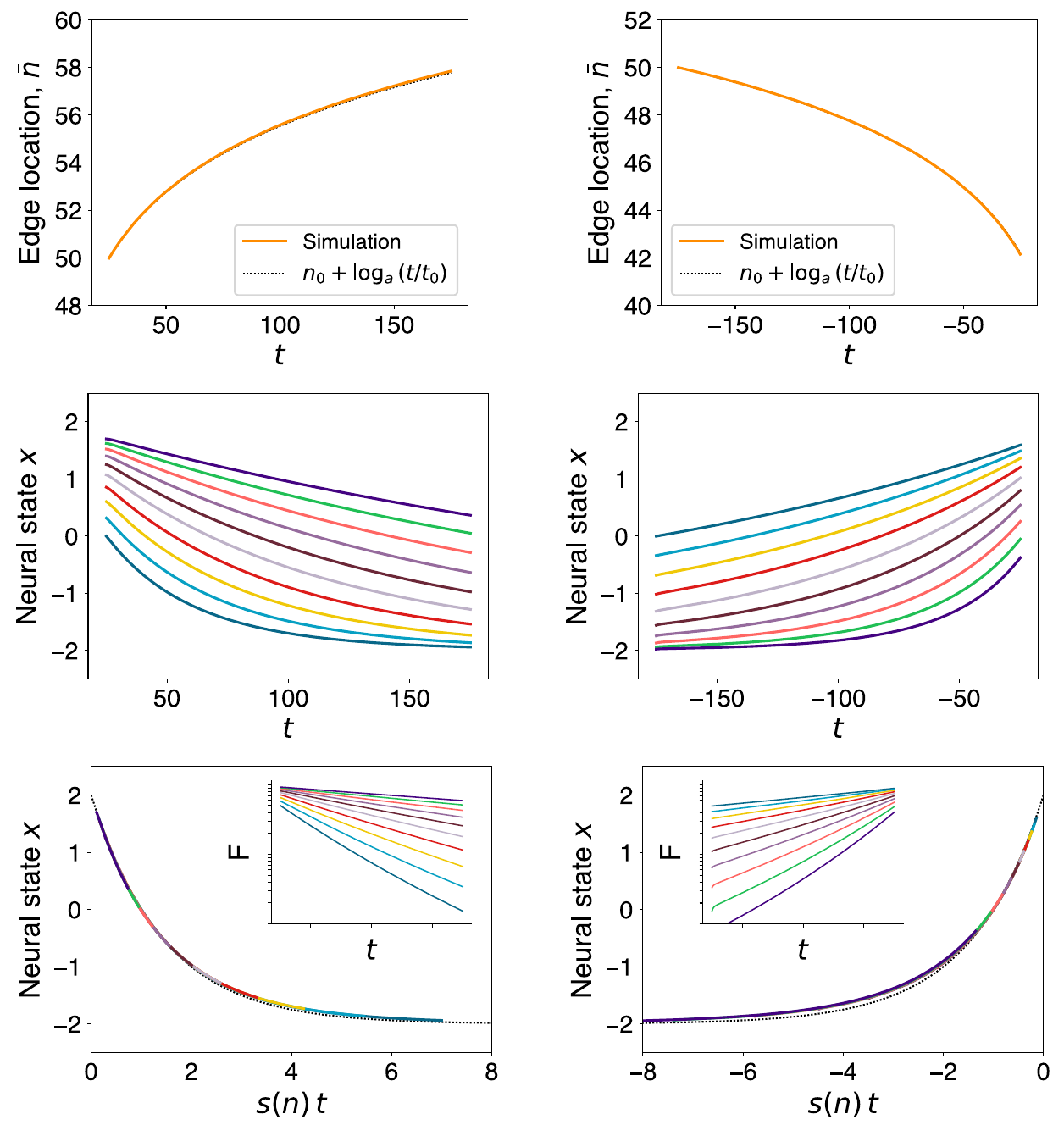}
    \caption{\textbf{Simulated dynamics of the continuous attractor neural network circuit model aligns with the
    theoretical properties of Laplace transform of the past and the future.}   
    Top: The location of the edge/bump as a function of $\tau$ moves as desired
    by the theory.
    Middle: Individual neurons in $\edgestate$ decay/ramp exponentially in time,
    as required for the Laplace transform of a delta function. 
    Bottom: Individual neurons in $\edgestate$  decay/ramp according to the same
    function. The time dynamics of individual neurons in $\edgestate$ differ
    only in their time scale.  In this figure, noise $\sigma_\xi^2 = 0$.
     \label{fig:asymmetric-time-scaling}
    }
\end{figure}

Rather than looking at the pattern of activity across the population at
different points in time, Figure~\ref{fig:asymmetric-time-scaling} describes 
activity of the edge network as a function of time.  
First, as can be seen from the top row, the location of the edge,
operationalized as the value of $n$ where activity passes through zero, moves
as desired (Equation~\ref{eq:barntime}) for both representations of the past (left) and the future (right).
This is a consequence of the choice for
$\feedback(n)$.
Second, following the time course of individual units in the edge network, we
find exponential firing as a function of time with a rate constant controlled
by $s(n)$. 
The bottom panel shows the time course of individual cells rescaled by each
cell's value of $s$.  It is clear that individual cells follow the same time
course up to a scaling factor.  This is a consequence of the logarithmic rate
at which the edge progesses. The inset plots the firing as a function of 
time on a logarithmic scale. The nearly straight lines are consistent with an exponential time
course, allowing a mapping between the activity of $\edgestate(n)$ and the real
Laplace transform.  Units participating in the representation of
the future (right) grow exponentially.

\begin{figure}
    \centering
    \includegraphics[width=0.8\columnwidth]{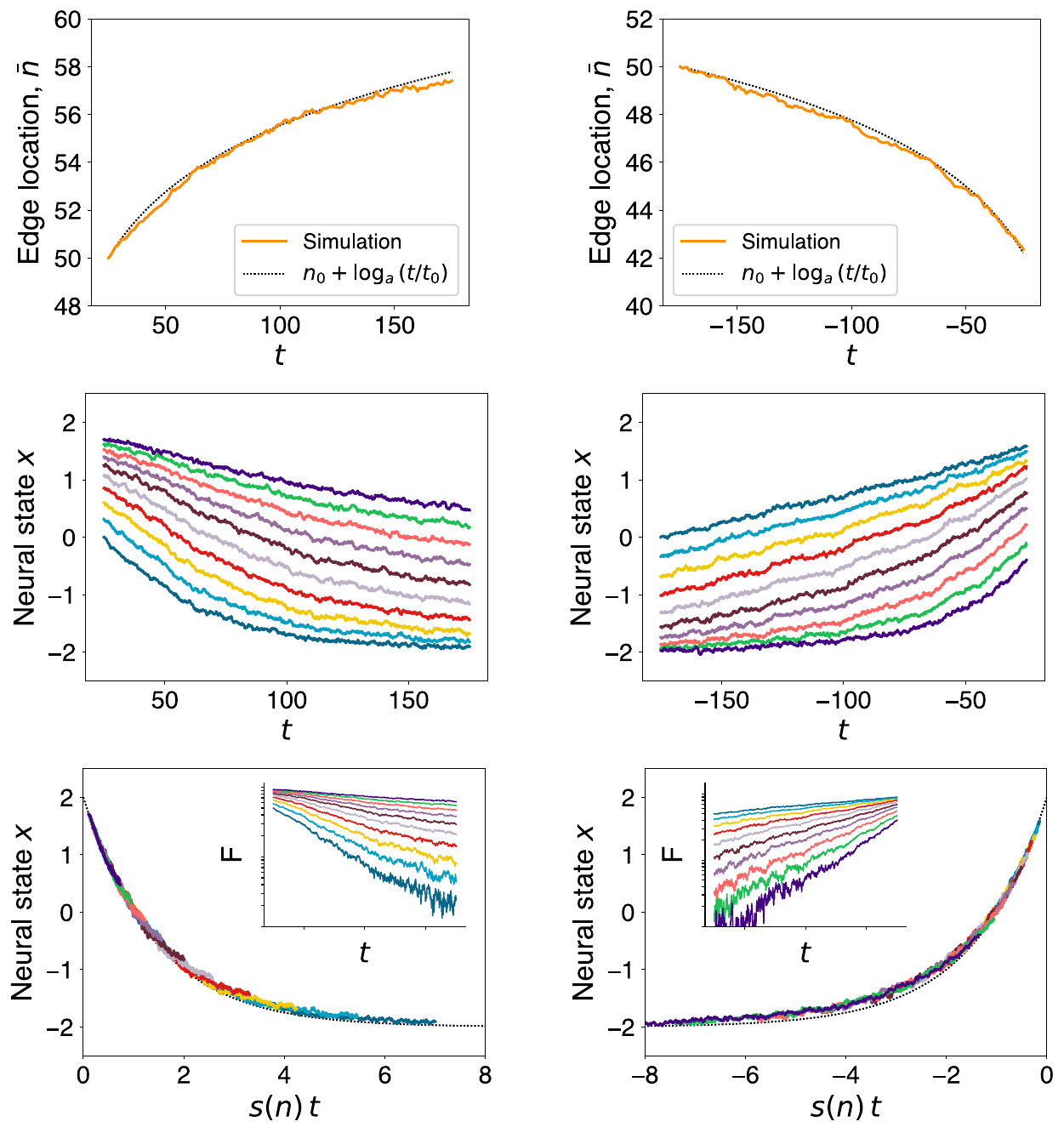}
    \caption{\textbf{
    Circuit model for Laplace transform is robust to noise.}   
    Layout analogous to Figure~\ref{fig:asymmetric-time-scaling}.  
    Here, the noise parameter $\sigma_\xi^2$ was increased from 0 to $10^{-3}$. 
    \label{fig:asymmetric-time-scaling-noise}
    }
\end{figure}

Because the representation of the time of future events requires exponential
growth,
tt would be impossible to construct an estimate of the Laplace transform of
the time of future events using independent leaky integrators
as in Equation~\ref{eq:leakyintegrator}.   With independent neurons, even a small
amount of noise would be rapidly amplified so that the pattern of activity 
can no longer reliably code for the time until the predicted event.
Figure~\ref{fig:asymmetric-time-scaling-noise} repeats the simulation in
Figure~\ref{fig:asymmetric-time-scaling} but in the presence of noise.  As can
be seen from visual inspection, the noise in the network is stable over time.
The location of the edge still moves as desired and the time course of
individual neurons still follows exponential functions albeit with
fluctuations around the mean.  Critically, neurons representing the time until
predicted future events grow smoothly even in the presence of noise.  
 
\begin{figure}
    \centering
    \includegraphics[width=0.6\columnwidth]{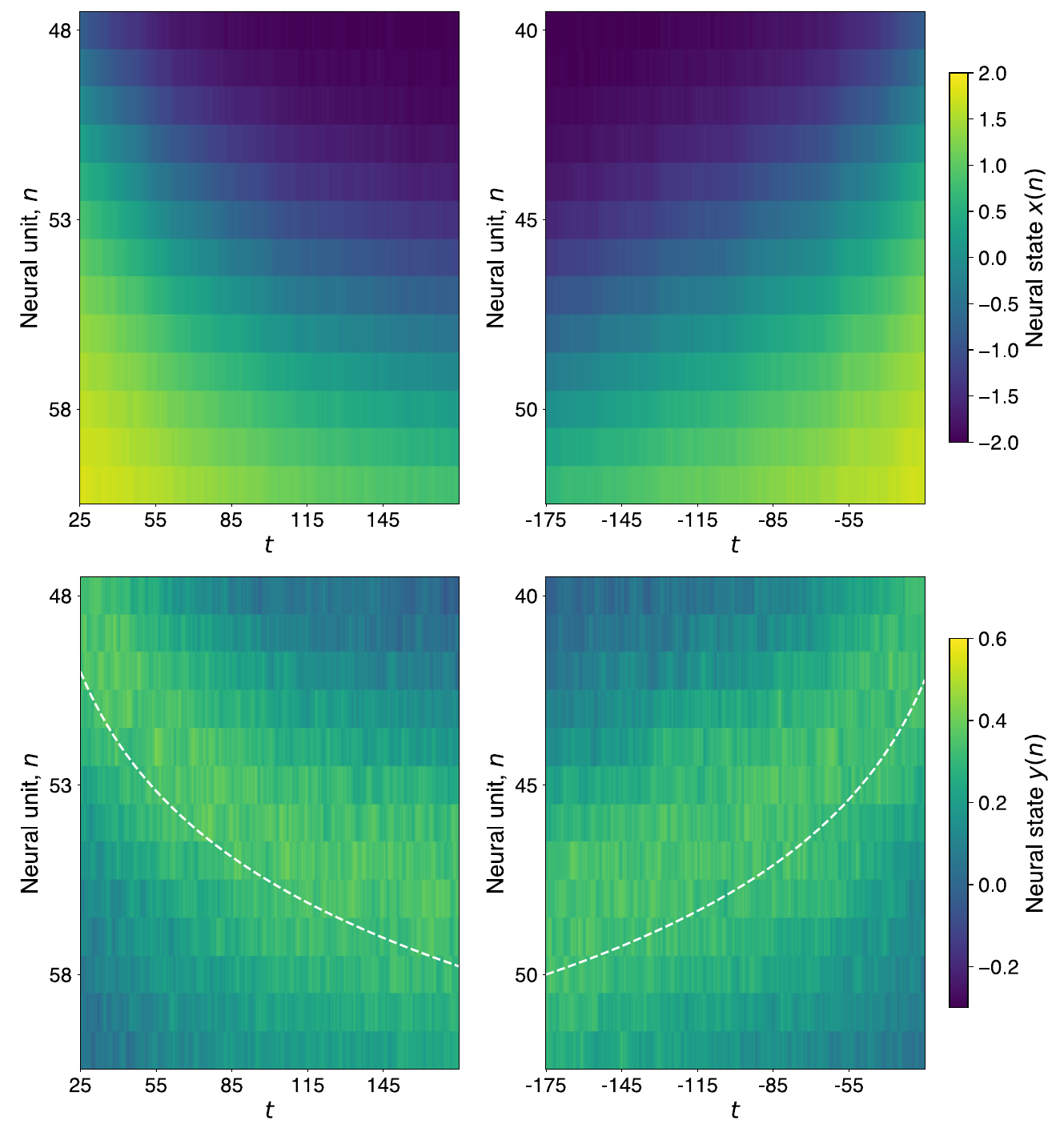}
    \caption{\textbf{Dynamics are similar to those observed in the brain.} 
    Layout is analogous to Figure~\ref{fig:DMmPFC}. 
    Top:
    The firing rate of each cell in an edge attractor network $\edgestate(n)$ as a
    function of time sorted on location in the network $n$.  On the left, the edge
    moves from small $n$ to large $n$ with $\feedback(n)$ positive.
    On the right, the edge moves from large $n$ to small $n$ with $\feedback(n)$ negative.
    The cells in these networks emulate neurons coding for the Laplace transform
    of the past and the Laplace transform of the future respectively.
    Bottom:  The activity of neurons in the bump network $\bumpstate(n)$ as a function of
    time for the corresponding edge attractors.  The neurons in these networks 
    emulate time cells coding for the past (left) and hypothesized ``future time cells'' (right).
    \label{fig:asymmetric-dynamics-noise}
    }
\end{figure}

The properties of the edge/bump networks in this paper are thus sufficient to
describe the basic properties of time-sensitive neurons observed in the brain. 
Figure~\ref{fig:asymmetric-dynamics-noise} plots the results of the same
simulation in Figure~\ref{fig:asymmetric-time-scaling-noise} using the same
conventions as used to show the neural data in Figure~\ref{fig:DMmPFC}.  In
this plot, units are sorted as a function of $n$ on the horizontal axis and the relative
firing rate of each neuron is shown \emph{via} color as a function of time.  Neurons in $\edgestate$ resemble the firing of exponentially decaying
and ramping neurons in the mammalian brain \citep{CaoEtal24}.  Neurons in
$\bumpstate$ coding for the past (lower left) fire sequentially with a
characteristic hook that resembles ``time cells'' observed in the hippocampus and
other brain regions \citep{CaoEtal22}.  The bump in the network associated
with the future (lower right) provides a mirror image of the time cells. These
hypothesized ``future time cells'' have more cells firing near the time of the
predicted event and with narrower temporal receptive fields.

\section{Discussion}

This paper developed a simple continuous attractor neural network with 
stable states that form an edge at any position along a line of neurons.
This attractor neural network is coupled to another set of units that express
a bump at a location that aligns with the edge.
The speed at which the edge moves is controlled by feedback from the bump
neurons to the edge neurons.  By allowing these connections to change as a
function of position along the line, the edge moves at different speeds.
Because we chose the velocity to be proportional to inverse time, the location of the edge/bump at any moment can be understood as the
logarithm of elapsed time.  Because the edge attractor has recurrent dynamics,
this network is robust to noise.  By changing the sign of the feedback signal we can
allow the edge/bump to move in either direction. 

The choices made in constructing the neural network were made to allow
a mapping of the population $\edgestate(n)$ onto 
 $\mathcal{L}\{f_t \}(s)$ where $f_t$ is a delta
function $t$ seconds in the past and $s(n) \propto a^{n}$.   This implies that
the bump is interpretable as an approximation of the original function $f$
projected onto log time. The bump is just $f$ convolved with some function
over log time describing the shape of the bump.  While the precise shape of
the connections within and between each layer affect the shape of this
function, the interpretation of $\bumpstate(n)$ as an approximation of the
original function projected on log time depends only on the
translation-equivariance of the stable states of $\edgestate(n)$ and the rate
at which the edge moves as a function of time.  This work allows a precise
connection between abstract cognitive models using a Laplace Neural Manifold
and circuit-level attractor neural network models. 

This framework is sufficiently general to represent both memory for the
past---delta functions $f$ that move further from zero as time passes---and
the future---delta functions $f$ that approach zero as time passes.
The function $\feedback(n)$ controls the speed at which the edge/bump moves
across the network.  To represent the future the edge moves in the opposite
direction and the sign of $\feedback(n)$ is reversed.  This results in units
in $\edgestate(n)$ that \emph{ramp upwards} exponentially in time with a
variety of time constants (Figure~\ref{fig:asymmetric-time-scaling}).  It would
be extremely challenging to build a network with this property with independent
units, as in Equation~\ref{eq:leakyintegrator}.   

\subsection{Limitations of this approach}

Our attractor network acts as a good approximator of a Laplace transform within a certain limited regime.  In particular, the approximation breaks down at small
times $|t - \tau|$, near the singularity at $t = \tau$, due to the network's inability to support arbitrarily large time derivatives.  
To a large extent, this is an unavoidable consequence of a 
neural implementation of the equations---there is no way to reach $\log 0$ in
a neural network. There are of course lower limits to the time scales at which
the nervous system operates.  For instance in the entorhinal cortex and
hippocampus, theta oscillations of 4-8~Hz set a lower limit to the time scale
that can be represented on a continuous timeline \citep{HassEtal02,Buzs05}. 
In our setup, we limit the magnitude of $\feedback(n)$ to
$\feedback_\mathrm{max}$, meaning that the edge velocity approaches a constant
at $t \rightarrow \tau$ instead of diverging as in
Eq.~\ref{eq:desired_velocity_vs_t}.  
This could correspond to a limitation in the ability to represent times in the very near past or future.  At the other extreme, the approximation also breaks down at large $|t - \tau|$ because noise overwhelms the signal of small time derivatives.  At large enough $|t - \tau|$, the edge would cease to move at the desired slow speed and would instead diffuse randomly.  This could correspond to an inability to represent times in the very distant past or future.

\subsection{Extensions of this approach} 
The present paper describes a minimal circuit to build the Laplace transform for
remembered past and predicted future times.  There are many possible elaborations
that would enhance its biological realism, computational stability, and
application to a wider range of cognitive problems.

\subsubsection{Symmetric weights; Alternatives to Laplace transform}
In a continuous attractor network, recurrent excitatory connections are typically chosen
to be symmetric as a function of distance within the network.
In this paper we chose the weights $\edgeedgeweight$ to give a shape of the edge 
that results in each unit's activation evolving in time as an exponential function.
This choice required weights that are not precisely symmetric  as a function of distance
within the network (Eq.~\ref{eq:edgeedgeweightresult}).
Different weights would have resulted in a different-shaped edge and the time
course of the individual neurons would no longer follow an exponential function.
Although the activity over that network would no longer be directly related to the Laplace
transform it would still be effective as a working memory representation over a
logarithmic timeline. 

The Appendix shows results analogous to
Figures~\ref{fig:asymmetric-translation}~and~\ref{fig:asymmetric-time-scaling}
computed with a network identical to the one in the main text except that the
connections $\edgeedgeweight$ are chosen to be a symmetric function of the
distance between units in $\edgestate$, as is more typical in continuous
attractor neural networks. The edge still moves in such a way that the location
of the edge $\bar{n}$ changes with $\log$ time.  This is a consequence of the
velocity controlled by $\feedback(n)$. With symmetric $\edgeedgeweight$, the
time courses of individual units are still rescaled versions of one another (mapping onto a single scaling function as in Eq.~\ref{eq:scaling-function}) with a time constant
that goes like $1/s_n$ (Figure~\ref{fig:asymmetric-time-scaling}~bottom).  This
means that the activity across $\edgestate(n)$ can still be understood as an
integral transform of the delta function $f$. 

The symmetric edge model described in the Appendix would be very difficult to
distinguish empirically from the model that maps onto the Laplace transform
described in the main text. The symmetric model results in $\bumpstate(n)$ being understandable
as a representation of the delta function $f$ as a function of log time, although
convolved with a slightly different bump shape.  
Cognitive models of retrieval time in working memory tasks rely on this logarithmic
timeline moreso than the shape of the bump of activity
\citep{HowaEtal15,TigaEtal21}. The distribution of time constants that would be
estimated from units in $\edgestate(n)$  with the symmetric model is identical
to the distribution estimated from the Laplace model.    

\subsubsection{Including recurrent bump dynamics}
A great deal of theoretical \citep{Amar77,Zhan96,RediTour97} and empirical
\citep{Taub98,KimEtal17} work has studied the properties of continuous bump
attractor neural networks for use in neuroscience and psychology
\citep{SchoSpen16}. In these networks, recurrent excitatory connections within a
population of neurons fall off as a function of distance within the network. In
addition, some form of inhibition, perhaps divisive normalization or global
inhibition, places an upper limit on the activity of the network. These two
constraints mean that the network can support stable states that take the form
of ``bumps'' of activity across the population. Because the strength of the
excitatory connections between neurons depends only on distance in the
network, the stable bump can be localized at effectively continuous locations. 

The activity over $\bumpstate(n)$ showed bumps of activity.  However, these did
not depend on recurrent connections between the units in $\bumpstate$ but were
inherited from input from the $\edgestate(n)$ \emph{via} connections
$\edgebumpweight$, which computes the derivative of $\edgestate(n)$ with respect to $n$. 
Inclusion of recurrent weights within $\bumpstate(n)$ would not have had a major impact on
the properties of this network.  We would expect such a network to be more resistant to noise
as recurrent weights  within both $\edgestate$ and $\bumpstate$ would tend to
stabilize the network. Because the bump attractor would be stable at all locations,
inputs from $\edgestate(n)$ that estimate the derivative should place the bump
attractor in the appropriate location. Although the shape of the bump would
certainly change depending on the form of the recurrent weights within units in
$\bumpstate$, as long as the location of the bump  coincides with the edge, 
activity over $\bumpstate(n)$ is interpretable as the original function
$f$  over log time convolved with some function describing the shape of the bump
attractor.

In this paper, we had the bump derived from input from the edge.
In principle, we could have chosen have the edge be derived from the bump.
In this case, we would have built a recurrent bump attractor network
with some mechanism to enable the location of the bump to move 
as a function of log time.
Indeed, others have implemented traveling waves in recurrent neural networks
with similar motivations \citep{KellEtal24}, although not in a way that
implements logarithmic scaling in time. 

\subsubsection{Dynamic control of edge/bump location}
The speed at which the edge/bump complex moves in this model is controlled by $\feedback(n)$.
The dynamics of this network require the stable states to be in an edge/bump
configuration.  Perturbations of network activity should relax to an edge/bump.
However, the location of the edge/bump should show cumulative error over time.
We would expect this error to be larger when the edge/bump is moving more slowly,
although it is possible that this effect can be mitigated by physiological mechanisms \citep{FranEtal06,YoshEtal08}.
In this way, there can be a mismatch between time as decoded
from the edge/bump location and
time as measured by an external clock.

The ability to move the edge/bump attractor allows this system in principle to 
be used for representing scalar quantities other than time.  
Choosing $\feedback(n)$ to be exponentially decreasing in $n$ allows the location of the
edge/bump to be interpretable as log time.
Suppose, however, that it were possible to modulate all of the connections by a time varying
external signal, something like $\feedback(n) \propto \alpha(t)$.  If $\alpha(t)=0$ the edge/bump would
stop moving, regardless of its location.  If $\alpha(t)>0$, then when it is
bigger the edge/bump moves faster.  If $\alpha(t)$ is controlled by the time
rate of change of some external variable such that $\alpha(t) = \frac{dz}{dt}$,
then the edge/bump moves like $\log_a z$.


This property is extremely useful in computational neuroscience.  For instance,
the place code in the hippocampus and related regions is believed to result from
integration of velocity signals. In the context of spatial navigation, neurons
participating in an edge attractor driven by spatial velocity would behave like
``border cells'' \citep{SolsEtal08,CampEtal18}. Critically, the edge attractor
model proposed here predicts that border cells should have a continuous
distribution of space constants controlling the width of their spatial receptive
fields. Cells participating in the bump attractor would behave in one dimension
as boundary vector cells \citep{BarrEtal06,LeveEtal09,SheeEtal21}; in two
dimensions conjunctions of boundary vector cells would have properties like
canonical hippocampal place cells \citep{BurgOKee96}.

Similarly, cognitive and neural models of decision-making accumulate
evidence over macroscopic periods of time \citep{Ratc78,GoldShad07}.  
At each moment, evidence accumulator models keep track of the distance in an abstract space
to decision bounds.  When the distance to one of the bounds reaches zero, a
behavioral decision is made. In this case, the velocity signal  $\alpha(t)$ is
the evidence available at each moment $t$ \citep{HowaEtal18}.  Neurons in $\edgestate(n)$
participating in evidence accumulation should behave more or less like classical
evidence accumulation cells ramping until a threshold when a decision is made
\citep{ShadNews01,HaneScha96}. Different units in $\edgestate(n)$ state ramp at
different rates depending on their value of $n$, exhibiting different ``evidence
constants.'' Neurons participating in the bump
during evidence accumulation should have receptive fields at different locations
along the evidence axis, not unlike observations in rodent cortex
\citep{MorcHarv16,KoayEtal22}. 

\subsection{Cognitive models using edge/bump attractors}

In many physical systems a coarse-grained description of the problem gives rise
to different physics than a finer-grained description.  In this sense the levels
of description each require their own models \citep{Ande72a}.  
Ideally, each of the levels should seamlessly connect to one another.
The major challenge of computational cognitive neuroscience is mapping neural
circuits onto cognitive models. Cognitive models provide a low-dimensional
description of behavior; if neural circuits can be mapped directly onto
cognitive models, then this provides a link from that circuit model to the
behavior of the organism.  For instance, the diffusion model \citep{Ratc78}
describes the degrees of freedom necessary to provide a more or less complete
description of behavior in many evidence accumulation tasks.  Insofar as that is
true, to the extent an RNN or an attractor model can be mapped onto the
diffusion model \citep{DaniEtal17} it can describe all the relevant behavior,
but may or may not accurately describe neural-level mechanisms. The flexibility
of Laplace/inverse representations for  cognitive models of a wide range of
behavioral tasks \citep{HowaEtal15,HowaEtal18,TigaEtal21} suggests that the
circuit developed here could be adapted for many cognitive operations.  The
Laplace Neural Manifold also provides an alternative hypothesis for constraining neural
mechanisms, which makes distinct predictions about the mapping onto neural data.

Most of contemporary computational neuroscience starts from neural circuits and attempts
to map onto behavior as an emergent property. 
The present approach---mapping cognitive models specified as Laplace transform
and inverse onto neural circuits---is conceptually distinct.  Rather than 
starting from neural circuits that seem biologically reasonable, this paper starts with 
a hypothesis for the collective behavior of neurons and then constructs a neural circuit
to satisfy those requirements.  Because the form of representation is motivated
by constraints from cognition, the mapping onto cognitive models is assured.

\subsubsection{Neural models for evidence accumulation}
The traditional way of interpreting the diffusion model is that each neuron in a
population provides a noisy estimate of the decision variable
\citep{ZandEtal14}.  This then suggests a dynamical system view where the two
options in the decision are associated with mutually exclusive attractors
\citep{Wang08,BogaEtal06}.  
Distance to bound can then be derived by a linear
projection  of the network state along the vector connecting the two attractors.
This approach makes several empirical predictions that are distinct
from the present edge/bump system.  Most notably, there is no natural way to map 
this idea onto the observation of decision-related sequences. 
Conversely, circuit models that create decision-related sequences do not provide an
account for why in other brain regions and/or behavioral tasks, monotonic
decision-related ramps are observed \citep{BrowEtal23}. 
Starting from the assumption that the decision variable is mapped onto a Laplace
Neural Manifold naturally generates both of these two functional cell types.  Neurons
in $\edgestate(n)$ should show classical ramping behavior as a function of the
decision variable; neurons in $\bumpstate(n)$ should show sequential activation
as a function of the decision variable.


Rather than voting for the time since an item was experienced---or the time
until a planned action is executed or a decision variable etc--- with their firing 
rates, in this view, neurons collectively represent information.  
This collective encoding differs from homogeneous voting when decomposing the unique and redundant contributions of individual neurons \citep{DaniEtAl16}.
Specifically, each neuron in
$\edgestate(n)$ spends most of its time in a high state of activation or a low
state of activation,  providing information about whether $\tau$ (or $T-\tau$)
is greater or lesser than $1/s_n$. Although the total activation across $\edgestate(n)$
is \emph{correlated} with $\log \tau$, more precise information about time can be read
off by observing the pattern across the entire population.  Similarly,
individual neurons in $\bumpstate(n)$ provide information about the number $\tau$ by 
means of the location of the bump; the average firing rate over the network is not 
correlated with $\tau$.  

The Laplace manifold mechanism also differs from fixed-point attractor models of
decision making in the origin of slow variables \citep{KhonFiet22,LangEtal23}.  Models that
represent a binary decision using two attractors must tune one collective
variable to a critical point in order to produce critical slowing down that
allows for evidence accumulation at a slower timescale than intrinsic neural
dynamics \citep{Wang02,DaniEtal17}.  The Laplace manifold mechanism instead has
built-in slow dynamics from the continuous symmetry of the attractor, which
requires tuning of all $N$ neurons.

\subsubsection{Data-independent operators for cognitive computation}
Finally, we note that properties of the Laplace transform make it well-suited to describe
data-independent operators for \emph{manipulating} information
\citep{HowaEtal15,HowaHass20}.   The network $\edgestate(n)$ represents the Laplace
transform of a delta function---essentially a single continuous number---on a
logarithmically-compressed number line.  There is a great deal of evidence that
the nonverbal number system is also logarithmically-compressed
\citep{DehaBran11,GallGelm00,NiedDeha09}.  If we could construct neural
operators to, say, add and subtract any arbitrary pair of numbers, this would be
extremely powerful for cognitive computation
\citep{FodoPyly88,GallKing11}.  If it were possible to combine information from two 
networks representing numbers with the representation used by $\edgestate(n)$, it would 
be straightforward to write out networks for addition and subtraction.
For instance, note that if we have two delta functions $f$ and $g$ centered at locations 
$a$ and $b$, then the convolution $f\ast g$ is centered at $a+b$.  Thus, adding two numbers
amounts to convolving the two delta functions. The Laplace transform
is extremely efficient for computing convolutions due to the fact that the transform 
of the convolution of two functions is simply the pointwise product of the two
transforms. In this way neural circuits for numerical cognition can be constructed by elaborating the edge/bump attractor
network developed here.

\section{Acknowledgments}

We gratefully acknowledge Sarah Marzen and Jim Crutchfield for organizing the
workshop ``Sensory Prediction, Engineered And Evolved'' at the Santa Fe
Institute where this work germinated, and all the participants
in that workshop for stimulating presentations and discussion.

\section{Code availability}

All code for simulating the continuous attractor network and creating the simulation figures can be found at \url{https://github.com/Collective-Logic-Lab/laplace-decisions}.

\bibliography{bibdesk}


\renewcommand{\thesection}{A\arabic{section}}
\renewcommand{\thetable}{A\arabic{table}}
\renewcommand{\thefigure}{A\arabic{figure}}
\renewcommand{\theequation}{A\arabic{equation}}
\setcounter{section}{0}
\setcounter{table}{0}
\setcounter{figure}{0}
\setcounter{equation}{0}

\clearpage

\section{Appendix}

\subsection{Using a symmetric edge}

Using a recurrent weight kernel $\edgeedgeweight(n-n')$ that is symmetric (even in $\Delta n = n - n'$) produces
symmetric edge shapes.  As discussed in the Methods, such a setup will still produce an edge that moves with the desired speed, with dynamics of individual neurons that are still simply rescaled in time.  Yet the symmetric case does not produce a representation of the Laplace transform, as the state of individual neurons does not depend exponentially on time.  

\begin{figure}
    \centering
    \includegraphics[width=0.8\columnwidth]{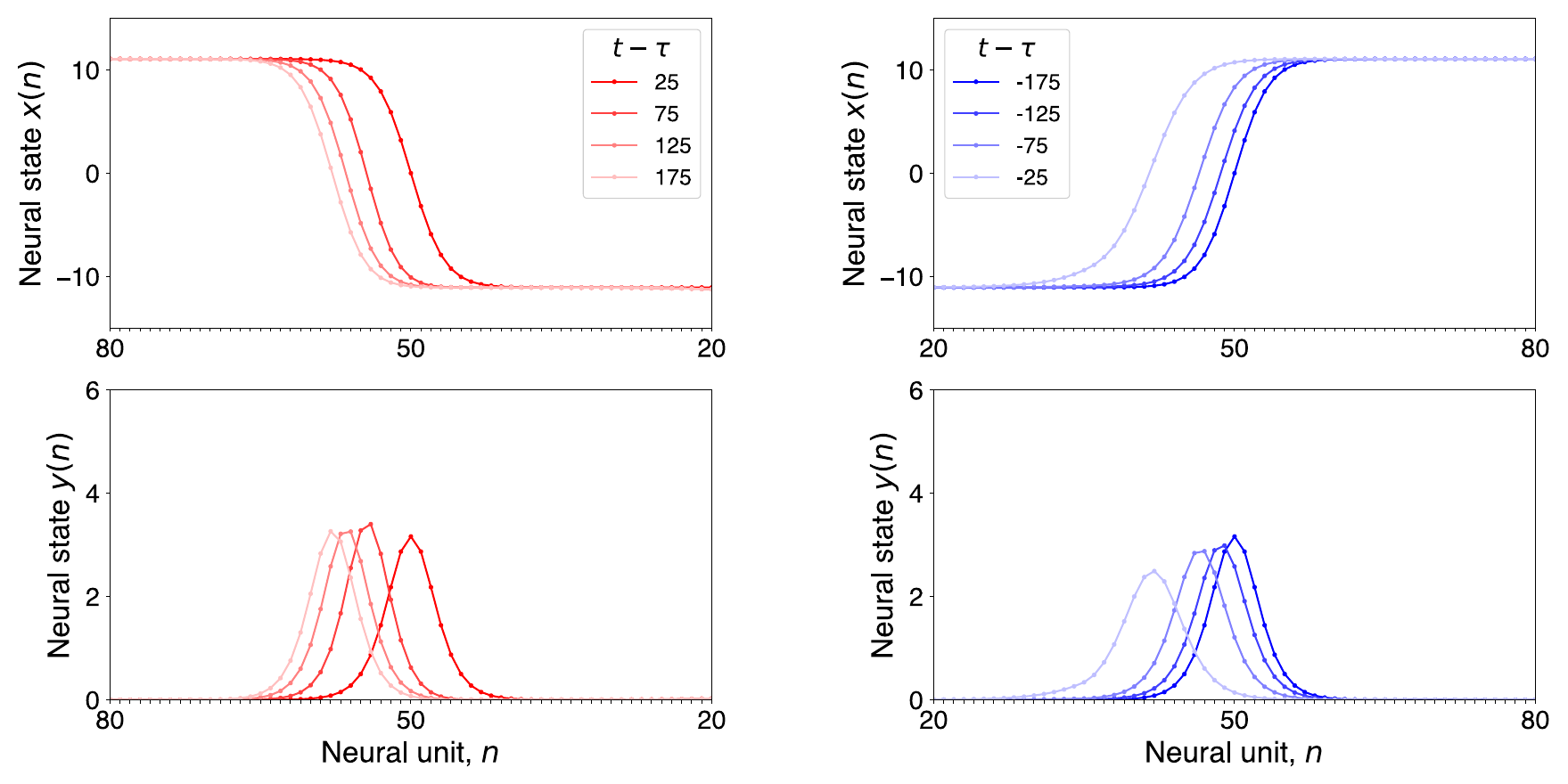}
    \caption{\textbf{
    The motion of symmetric edges.}   
    Layout analogous to Figure~\ref{fig:asymmetric-translation}.  
    However, this uses the symmetric recurrent weight kernel in Eq.~\ref{eq:symmetric-kernel} instead of Eq.~\ref{eq:asymmetric-kernel}. 
    \label{fig:symmetric-translation}
    }
\end{figure}

\begin{figure}
    \centering
    \includegraphics[width=0.8\columnwidth]{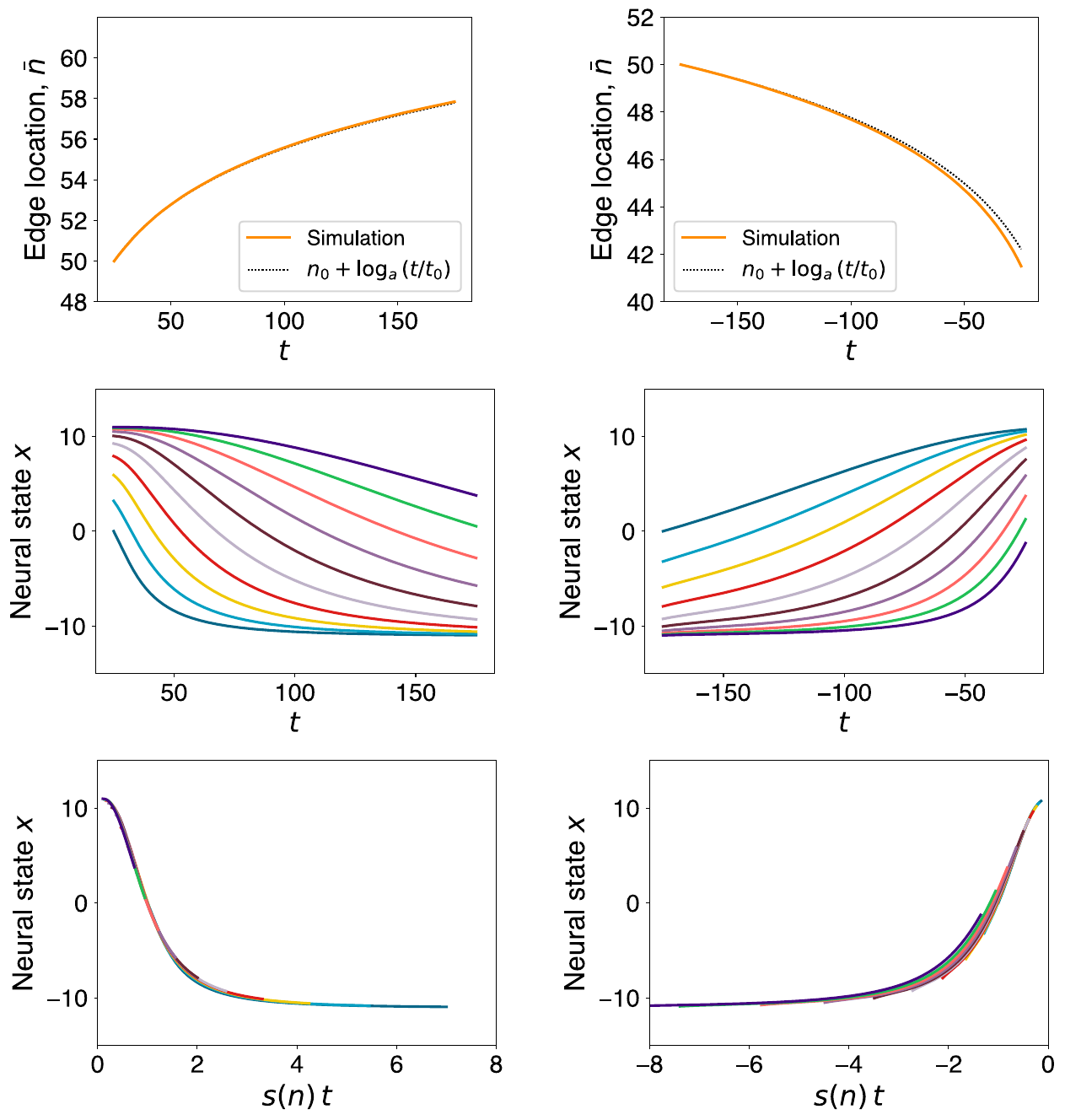}
    \caption{\textbf{
    Symmetric case reproduces time scaling properties except for exponential decay.}   
    Layout analogous to Figure~\ref{fig:asymmetric-time-scaling}.  
    However, this uses the symmetric recurrent weight kernel in Eq.~\ref{eq:symmetric-kernel} instead of Eq.~\ref{eq:asymmetric-kernel}. 
    \label{fig:symmetric-time-scaling}
    }
\end{figure}

To demonstrate this case explicitly, we simulate the dynamics using a Gaussian kernel:
\begin{equation}
    \edgeedgeweight(\Delta n) = J \exp(-\Delta n/\sigma_k^2).
    \label{eq:symmetric-kernel}
\end{equation}
The resulting dynamics are shown in Figs.~\ref{fig:symmetric-translation} and \ref{fig:symmetric-time-scaling}, using $J=22$, $\sigma_k = 1$ and other parameters as in Table~\ref{tab:parameters}.

\end{document}